\documentclass[twocolumn,showpacs,preprintnumbers,amsmath,amssymb]{revtex4}

\usepackage{graphicx}
\usepackage{dcolumn}
\usepackage{bm}

\begin{document}

\title{Dynamics of the universe in the modified unimodular theory of gravity}
\author{Robert D. Bock\footnote{email: robert.bock@pra-corp.com}}

\affiliation{Propagation Research Associates, Inc., 1275 Kennestone Circle, Suite 100, Marietta, GA, 30066}

\date{\today}

\begin{abstract}
The equations that govern the dynamics of the universe in the modified unimodular theory of gravity are derived.  We find a mechanism for inflation in the early universe without postulating a false vacuum state during the first $10^{-35}$ seconds after the Big Bang.  In addition, we find a natural explanation for the acceleration of the universe without resorting to dark energy.
\end{abstract}

\maketitle
The Friedmann-Lema\^itre $\Lambda\text{CDM}$ model of cosmology has been accepted by the scientific community as the new Standard Model of Cosmology \cite{steinhardtostriker,bahcalletal}.  It supercedes the previous Standard Model of Cosmology, embracing all of its accomplishments, and claims additional success.  This model agrees closely with a wide range of observations, including measurements of the abundance of primordial elements, CMB anisotropies, the age of the universe, the luminosity of supernovae, and the large scale structure of the universe.  According to the $\Lambda\text{CDM}$ model, the universe is spatially flat and was initiated with the Big Bang, a state of infinite density and temperature, approximately $15 \times 10^9$ years ago.  This was followed by a potential, or vacuum, energy-dominated (inflation) phase, a radiation-dominated phase, and a matter-dominated phase.  It is believed that the universe is presently transitioning from the matter-dominated phase to a cosmological constant-dominated phase.  

First reported in Refs. \cite{perlmutteretal, riess1, riess2}, observations of Type Ia Supernova (SN Ia) indicate that the universe is accelerating.  The $\Lambda\text{CDM}$ model attributes this acceleration to the cosmological constant, $\Lambda$, which was originally introduced into general relativity by Einstein \cite{einstein1917} in order to permit homogeneous, static solutions of the field equations.  However, the introduction of the cosmological constant brings a number of problems in its wake, including the well known cosmological constant, or fine-tuning, problem.  This results from the observation that the contribution to the vacuum energy density from quantum fields behaves like a cosmological constant, and is according to modern particle theories orders of magnitude larger than the measured cosmological constant, which is crudely approximated by $\Lambda \approx H_0^2$, where $H_0$ is the present value of the Hubble parameter.  Consequently, considerable effort is being exerted to replace $\Lambda$ in the $\Lambda\text{CDM}$ model with more general forms of dark energy that are typically described by scalar fields such as quintessence, K-essence, tachyon, phantom and dilatonic models (see \cite{copelandetal} for an excellent review).  

Recently, a modified unimodular theory of gravity was introduced \cite{bock2} that replaces the covariant divergence law with a modified divergence law based on the metric decomposition of unimodular relativity.  It was shown that the resulting equations of motion acquired an additional term in the Newtonian limit that could account for the anomalous accelerations observed on the galactic scale without the need for dark matter.  In the following, we examine the consequences of this modified divergence law on the dynamical equations of the universe and find a natural explanation for the acceleration of the universe without resorting to dark energy. 

According to the unimodular theory of gravity \cite{andersonfinkelstein} (see also \cite{buchmullerdragon,ngdam,weinberg}), the metric tensor is reducible under the general coordinate category into two nontrivial geometric objects: $g$, the determinant of the metric tensor, and $\gamma_{\mu\nu}$, the relative tensor $g_{\mu\nu}/(\sqrt{-g})^{1/2}$ of determinant $-1$.  The determinant determines entirely the measure structure of space–time, while the relative tensor alone determines the null-cone or causal structure.  Unimodular relativity assumes a background measure field $\sqrt{-g}=\sigma(x^\alpha)$ so that:
\begin{equation}
\label{metric}
g_{\mu\nu}=\sigma(x^\alpha)^{1/2}\gamma_{\mu\nu},
\end{equation}
and satisifies this condition in the action with the method of Lagrange undetermined multipliers.  The field equations are given by the traceless part of Einstein's field equations:                      
\begin{equation}
\label{traceless}
R^{\mu}_{\nu}-\frac{1}{4}R\delta^{\mu}_{\nu}=8\pi G\left(T^{\mu}_{\nu}-\frac{1}{4} T\delta^{\mu}_{\nu}\right),
\end{equation}
where we have assumed $c=1$.  The covariant divergence of the above equations yields:
\begin{equation}
\label{firstidentity}
8\pi G T^{\mu}_{\nu;\mu}=\frac{1}{4}\delta^{\mu}_{\nu}\left(R+ 8\pi G T \right)_{,\mu}.
\end{equation}
Assuming the standard covariant divergence law:
\begin{equation}
\label{divergencelaw}
T^{\mu}_{\nu;\mu}=0,
\end{equation}
the cosmological constant emerges as a constant of integration in the field equations.  This formulation has the attractive property that the contribution of vacuum fluctuations automatically cancels on the right hand side of Eq. (\ref{traceless}) \cite{weinberg}.  In Ref. \cite{bock2}, the covariant divergence law (\ref{divergencelaw}) was replaced with:
\begin{equation}
\label{modconservation2b}
T^{\mu}_{\nu;\mu}=\frac{\sigma_{,\mu}}{\sigma}
\left[AT^{\mu}_{\nu}+B T \delta^{\mu}_{\nu}\right] + C T^{\alpha\mu}g_{\sigma\nu}\Gamma^{\prime\sigma}_{\alpha\mu},
\end{equation}
where $A$, $B$, and $C$ are arbitrary constants and
\begin{equation}
\label{connection2defs}
\Gamma^{\prime\mu}_{\alpha\beta} \equiv  \frac{\gamma^{\mu\lambda}}{2}\left[\gamma_{\alpha \lambda ,\beta} + \gamma_{\beta \lambda ,\alpha} - \gamma_{\alpha \beta ,\lambda}\right].
\end{equation}
Eqs. (\ref{traceless}), (\ref{firstidentity}), and (\ref{modconservation2b}) may be used to determine the dynamics of the universe, given a source term describing the energy content.  In the following we set $C=0$, since otherwise the results depend on the spatial coordinate system chosen. 
                         
The Friedmann-Robertson-Walker (FRW) metric is \cite{weinbergbook}:
\begin{equation}
\label{frwmetric}
ds^2=-dt^2+a^2(t)\left[\frac{dr^2}{1-Kr^2}+r^2\left(d\theta^2+\sin^2\theta d\phi^2 \right)          \right],
\end{equation}
where $a(t)$ is the scale factor, $K$ is the curvature constant, $t$ is cosmic time, and the coordinates ${r,\theta,\phi}$ are the comoving coordinates of a free particle.  The measure field of the FRW metric is:
\begin{equation}
\label{sigma}
\sigma\equiv\sqrt{-g}=\frac{a^3r^2\sin\theta}{\sqrt{1-Kr^2}}.
\end{equation}
The relevant curvature tensor components are \cite{kolbandturner}:
\begin{eqnarray}
\label{components}
R^{0}_{0}&=&\frac{3\ddot{a}}{a}  \\
R^{i}_{j}&=&\left(\frac{\ddot{a}}{a}+\frac{2\dot{a}^2}{a^2}+\frac{2K}{a^2}\right)\delta^{i}_{j}  \\
R&=&6\left(\frac{\ddot{a}}{a}+\frac{\dot{a}^2}{a^2}+\frac{K}{a^2} \right),
\end{eqnarray}
where a dot denotes differentiation with respect to $t$ and $i,j=1\ldots 3$.  We assume an ideal perfect fluid as the source of the energy momentum tensor $T^{\mu}_{\nu}$:
\begin{equation}
\label{idealfluid}
T^{\mu}_{\nu}=\text{Diag}(-\rho,p,p,p),
\end{equation}
where $\rho$ and $p$ are the energy density and the pressure density of the fluid, respectively.  Substituting the ideal fluid energy-momentum tensor into Eq. (\ref{traceless}) we find only one independent equation (as opposed to two independent equations in the standard case):
\begin{equation}
\label{friedmann1}
\dot{H}=\frac{\ddot{a}}{a} - \frac{\dot{a}^2}{a^2}=-4\pi G\left(\rho+p \right)+\frac{K}{a^2},
\end{equation}  
where $H\equiv \frac{\dot{a}}{a}$ is the Hubble parameter.  Another equation follows by combining Eqs. (\ref{firstidentity}) and (\ref{modconservation2b}):
\begin{eqnarray}
\label{friedmann2}
\frac{\dddot{a}}{a} + \frac{\ddot{a}\dot{a}}{a^2} - \frac{2\dot{a}^3}{a^3} - \frac{2K\dot{a}}{a^3} + \frac{4\pi G}{3}\left(3\dot{p}-\dot{\rho} \right) \\ \nonumber
=\frac{16\pi G \dot{a}}{a}\left[3Bp-\rho(A+B) \right].
\end{eqnarray}
From the modified divergence law (\ref{modconservation2b}) we also find:
\begin{equation}
\label{friedmann3}
\dot{\rho}+ 3\frac{\dot{a}}{a}\left(\rho+p\right)=3\frac{\dot{a}}{a}\left[3Bp-\rho(A+B)\right].
\end{equation}
For $A=B=0$ this produces the standard continuity equation.  For non-zero $A$ and $B$ this reduces to the standard continuity equation when:
\begin{equation}
\label{ABcondition}
1+\frac{A}{B}=\frac{3p}{\rho}.
\end{equation}
Eqs. (\ref{friedmann1}), (\ref{friedmann2}), and (\ref{friedmann3}) determine the dynamics of the universe for an ideal perfect fluid source.  As in the standard case, only two of these equations are independent equations.  It is convenient to combine Eqs. (\ref{friedmann1}) and (\ref{friedmann2}).  Solving for $\frac{\dot{a}}{a}$ in Eq. (\ref{friedmann1}) and substituting into Eq. (\ref{friedmann2}) we find:
\begin{eqnarray}
\label{friedmann4}
&&\frac{\ddot{a}}{a}-\frac{\dddot{a}}{\dot{a}}=16\pi G\times\\ \nonumber
&&\left[\frac{a}{\dot{a}}\left(\frac{\dot{p}}{4}-\frac{\dot{\rho}}{12} \right) -p\left(3B+\frac{1}{2}\right)+\rho\left(A+B-\frac{1}{2}\right)            \right].
\end{eqnarray}
Another useful relationship follows by solving for $\frac{\ddot{a}}{a}$ in Eq. (\ref{friedmann1}) and substituting into Eq. (\ref{friedmann2}):
\begin{eqnarray}
\label{friedmann4b}
H^2 &=& \frac{\dddot{a}}{\dot{a}} -\frac{K}{a^2}+\frac{4\pi G}{3H}\left[3\dot{p}-\dot{\rho}  \right]    \\ \nonumber
&+&16\pi G\left[\rho\left(A+B-\frac{1}{4} \right)-p\left(3B+\frac{1}{4}\right)  \right].
\end{eqnarray}

We now examine the dynamics of the universe for the radiation-dominated and matter-dominated epochs of its evolution.  The equations of state for radiation and dust are:
\begin{equation}
\label{radiation}
\text{Radiation:}\;\; p=\frac{\rho}{3} 
\end{equation}
\begin{equation}
\label{dust}
\text{Dust:}\;\; p=0.
\end{equation}
Substituting the above equations of state into Eq. (\ref{friedmann3}) we find:
\begin{equation}
\label{radcontinuity}
\text{Radiation:}\;\; \dot{\rho}=-3\frac{\dot{a}}{a}\rho\left(A+\frac{4}{3}\right) 
\end{equation}
\begin{equation}
\label{mattercontinuity} 
\text{Dust:}\;\;  \dot{\rho}=-3\frac{\dot{a}}{a}\rho(A+B+1).
\end{equation}
Solving the above differential equations yields:
\begin{equation}
\label{solutionradcontinuity}
\text{Radiation:}\;\; \rho \propto a^{-3\left(A+\frac{4}{3}  \right)}
\end{equation}
\begin{equation}
\label{solutionmattercontinuity} 
\text{Dust:}\;\; \rho \propto a^{-3\left(A+B+1 \right)}.
\end{equation}
We see that energy creation occurs in a radiation- (matter-) dominated universe for $A<-\frac{4}{3}$ ($A+B<-1$), assuming $\frac{\dot{a}}{a}>0$.  Hence, Eqs. (\ref{radcontinuity}) and (\ref{mattercontinuity}) provide a mechanism for the generation of energy in the universe.  Note that this energy creation is exponential if $H=\frac{\dot{a}}{a}=\text{constant}$. 

Using Eqs. (\ref{friedmann3}) and (\ref{friedmann4}) we find for radiation- and matter-dominated universes:
\begin{equation}
\label{expradiationeq}
\text{Radiation:}\;\;\frac{\ddot{a}}{a}-\frac{\dddot{a}}{\dot{a}}=16\pi G\rho\left(A-\frac{2}{3}    \right)
\end{equation}
\begin{equation}
\label{expdusteq}
\text{Dust:}\;\;\frac{\ddot{a}}{a}-\frac{\dddot{a}}{\dot{a}}=4\pi G\rho\left(5A+5B-1     \right).
\end{equation}
We see that the dynamical equations provide a mechanism for accelerated expansion in both radiation-dominated and matter-dominated cosmologies.  In particular, in a matter-dominated universe accelerated expansion is possible without dark energy.  

Using Eqs. (\ref{friedmann4b}), (\ref{radcontinuity}), and (\ref{mattercontinuity}) we find:
\begin{equation}
\label{H2rad}
\text{Radiation:}\;\;\frac{K}{H^2a^2}=\frac{\dddot{a}}{H^2\dot{a}}+\frac{16\pi G\rho}{H^2}\left(A-\frac{1}{3}\right)-1
\end{equation}
\begin{equation}
\label{H2dust}
\text{Dust:}\;\;\frac{K}{H^2a^2}=\frac{\dddot{a}}{H^2\dot{a}}+\frac{20\pi G\rho}{H^2}\left(A+B\right)-1.
\end{equation}
These equations describe the evolution of the curvature of the universe in radiation-dominated and dust-dominated cosmologies, respectively.  Note that they depend on the coefficients $A$ and $B$ and the third derivative of the scale factor, in addition to the energy density.

The dynamical equations for a radiation-dominated (dust-dominated) universe have a simple solution for the case $A=\frac{2}{3}$ ($A+B=\frac{1}{5}$).  In this case Eqs. (\ref{expradiationeq}) and (\ref{expdusteq}) become:
\begin{equation}
\label{expradiationeq2}
\frac{\ddot{a}}{a}-\frac{\dddot{a}}{\dot{a}}=0,
\end{equation}
which has the solution:
\begin{equation}
\label{exponentialexpansion}
a\propto \exp(H t),
\end{equation}
where $H=\text{constant}$.  Thus, the dynamical equations provide a mechanism for inflation during the early radiation-dominated phase.  Consequently, it is not necessary to postulate a previous false-vacuum state to drive the inflation.  If we consider the evolution of the curvature for the radiation- and matter-dominated universes undergoing the accelerated expansion (\ref{exponentialexpansion}) we find:
\begin{equation}
\label{curvaturetozero}
\frac{K}{H^2a^2} \sim  \rho\sim\exp(-\epsilon_0 Ht),
\end{equation}
where $\epsilon_0 = 6$ ($\epsilon_0 = \frac{18}{5}$) for radiation (dust).  Therefore, the curvature $K$ goes to zero as $t \rightarrow \infty$, which also results from standard models of inflation \cite{blauandguth}.  

Another simple solution follows for the case of constant energy density.  According to Eq. (\ref{radcontinuity}) (Eq. (\ref{mattercontinuity})), $\rho$ is constant when $A=-4/3$ ($A+B=-1$) for radiation (dust).  Therefore, Eqs. (\ref{expradiationeq}) and (\ref{expdusteq}) may be written:
\begin{equation}
\label{radb}
\frac{\ddot{a}}{a}-\frac{\dddot{a}}{\dot{a}}=-\omega_0^2,
\end{equation}
where $\omega_0^2 = 32\pi G\rho$ ($\omega_0^2 = 24\pi G\rho$) for radiation (dust) of constant energy density.
Solving the above differential equation, we find an inflationary solution with a quadratic time in the exponential:
\begin{equation}
\label{exponentialexpansion2}
a\propto \exp \left(\frac{\omega_0^2 t^2}{4}\right).
\end{equation}
In this case, $H$ is not constant:
\begin{equation}
\label{Hquadraticexponential}
H\propto\left(\frac{\omega^2_0}{2} \right)t.
\end{equation}
The curvature for the radiation- and matter-dominated universes undergoing the accelerated expansion (\ref{exponentialexpansion2}) obeys:
\begin{equation}
\label{curvaturetozero2}
\frac{K}{H^2a^2} \sim \frac{\dddot{a}}{H^2\dot{a}} + \frac{1}{H^2} -1\sim\frac{1}{t^2}.
\end{equation}
Again, we find $K \rightarrow 0$ as $t \rightarrow \infty$.

In summary, unimodular relativity with the modified divergence law (\ref{modconservation2b}) clarifies a number of fundamental issues in modern cosmology.  The cosmological constant does not appear in the formalism for non-zero $A$, $B$ and the contribution of vacuum fluctuations is canceled out in the field equations.  In addition, it provides a mechanism for the acceleration of the present dust-dominated universe without dark energy.  Consequently, there is no need to assume that we are coincidentally living during the transition from a matter dominated universe to a $\Lambda$-dominated universe.  Moreover, it provides a mechanism for inflation in the early universe without postulating a false-vacuum state during the first $10^{-35}$ seconds after the Big Bang.  As a result, many of the consequences of standard inflation models remain valid.  Furthermore, one may describe the history of the universe with only two epochs, a radiation-dominated epoch followed by a matter-dominated epoch, thus removing the vacuum-dominated (inflation) phase and the $\Lambda$-dominated phase from the description of the early and late history of the universe.  Note that it is not necessary to postulate that all of the energy content of the universe was present at the Big Bang (as must be done in the $\Lambda\text{CDM}$ model if its history were to be described only by radiation-dominated and matter-dominated epochs) since the dynamical equations provide a mechanism for energy creation.

While the quantities $A$ and $B$ were introduced in the modified divergence law (\ref{modconservation2b}) as constants, future work may require a time variation ($A(t)$ and $B(t)$) to describe the various stages of the evolution of the universe.  Thus, it is possible that there is an epoch in the history of the universe during which the standard covariant conservation law (\ref{divergencelaw}) is satisfied.

\bibliography{de}
\end{document}